\input harvmac
\input psfig
\newcount\figno
\figno=0
\def\fig#1#2#3{
\par\begingroup\parindent=0pt\leftskip=1cm\rightskip=1cm\parindent=0pt
\global\advance\figno by 1
\midinsert
\epsfxsize=#3
\centerline{\epsfbox{#2}}
\vskip 12pt
{\bf Fig. \the\figno:} #1\par
\endinsert\endgroup\par
}
\def\figlabel#1{\xdef#1{\the\figno}}
\def\encadremath#1{\vbox{\hrule\hbox{\vrule\kern8pt\vbox{\kern8pt
\hbox{$\displaystyle #1$}\kern8pt}
\kern8pt\vrule}\hrule}}
\def\underarrow#1{\vbox{\ialign{##\crcr$\hfil\displaystyle
 {#1}\hfil$\crcr\noalign{\kern1pt\nointerlineskip}$\longrightarrow$\crcr}}}
%
\overfullrule=0pt

%
\def\tilde{\widetilde}
\def\bar{\overline}

\def\R{{\bf R}}

\font\zfont = cmss10 

\def\bigone{\hbox{1\kern -.23em {\rm l}}}
\def\ZZ{\hbox{\zfont Z\kern-.4emZ}}

\Title{hep-th/0306083}
{\vbox{\centerline{A Note On The Chern-Simons}
\bigskip
\centerline{And Kodama Wavefunctions }}}
\smallskip
\centerline{Edward Witten}
\smallskip
\centerline{\it Institute For Advanced Study, Princeton NJ 08540 USA}


\medskip

\noindent Yang-Mills theory in four dimensions formally admits an
exact Chern-Simons wavefunction.  It is an eigenfunction of the
quantum Hamiltonian with zero energy.  It is known to be
unphysical for a variety of reasons, but it is still interesting
to understand what  it describes.  We show that in expanding
around this state, positive helicity gauge bosons have positive
energy and negative helicity ones have negative energy. Some of
the negative energy states have negative norm.  We also show that
the Chern-Simons state is the supersymmetric partner of the naive
fermion vacuum in which one does not fill the fermi sea. Finally,
we give a sort of explanation of ``why'' this state exists.
Similar properties can be expected for the analogous Kodama
wavefunction of gravity.

 \Date{June, 2003}
Four-dimensional Yang-Mills theory has the surprising property of
admitting an exact zero energy eigenfunction of the Schrodinger
equation, the wave-function being the exponential of the
Chern-Simons form. This wavefunction, which has been known for a
long time (the only original reference I know of is \ref\jackiw{R.
Jackiw, ``Topological Investigations In Quantized Gauge
Theories,'' p. 258, exercise 3.7, in S. B. Treiman et. al. {\it
Current Algebra And Anomalies} (World Scientific, 1985).}, where
it is presented as an exercise!), is highly unnormalizable. It is
constructed without the asympototic freedom and coupling constant
renormalization that are needed for the standard quantization of
Yang-Mills theory, which makes contact both with nonperturbative
lattice calculations and with real strong interaction and weak
interaction experiments. Finally, it is not invariant under CPT
(as we discuss more fully later), so on general grounds it could
not be the ground state of a quantum field theory.   For all these
reasons and more we will find later, the Chern-Simons wave
function of Yang-Mills theory is not the physical ground state of
the theory. Nonetheless, one would like to know how it should be
interpreted, and in some sense, ``why'' it exists.  Answering
these questions will be the goal of the present paper.

It is easy to describe directly the Chern-Simons wavefunction of
Yang-Mills theory. The Hamiltonian of Yang-Mills theory is
\eqn\turno{H={1\over 2 g^2}\int d^3x\Tr(E^2+B^2)={1\over 2}\int
d^3x\Tr\left(-g^2{\delta^2\over \delta A(x)^2}+ {1\over
g^2}B^2\right).} Here $g$ is the gauge coupling, and $E_i=F_{0i}$
and $B_i={1\over 2}\epsilon_{ijk}F_{jk}$ are the electric and
magnetic fields.  The canonical momentum is $\Pi= E/g^2$, and
quantum mechanically it becomes $-i\delta/\delta A$, whence the
second formula in \turno.  Given the expression for the
Hamiltonian, it is clear that any wavefunction $\Psi$ with
\eqn\urno{0=(E+iB)\Psi=i\left(-{ g^2}{\delta\over \delta
A}+B\right)\Psi} is also an eigenfunction of the Hamiltonian with
$H\Psi=0$.  Moreover, if $I$ is the Chern-Simons functional,
$I={1\over 4\pi}\int d^3x\epsilon^{ijk}\left(A_i\partial_j
A_k+{2\over 3}A_iA_jA_k\right)$, then $\delta I/\delta A=B/2\pi$,
so that \eqn\toffo{\Psi= \exp\left((2\pi/g^2)I(A)\right)}
 obeys
\urno\ and hence is an eigenstate of the Hamiltonian with zero
energy. This is what we call the Chern-Simons state.  It is far
from being normalizable, since $I(A)$ has no properties of
positivity -- it changes sign under parity. It would be equally
good to have a state annihilated by $E-iB$, and clearly
$\tilde\Psi= \exp(-(2\pi/g^2)I(A))$ does this job.\foot{In the
nonabelian case, $\Psi$ and $\tilde \Psi$ are not invariant under
homotopically non-trivial gauge transformations. We ignore this.
Along with the unnormalizability, lack of CPT invariance, etc.,
and additional properties that we will see below, this is one more
reason that the Chern-Simons state is formal and does not really
correspond to a sensible physical theory.}

The Chern-Simons wavefunction of Yang-Mills theory has an even
more surprising gravitational analog, commonly called the Kodama
state \ref\kodama{H. Kodama, ``Holomorphic Wave Function Of The
Universe,'' Phys. Rev. {\bf D42} (1990) 2548.}.  Some authors have
proposed the Kodama wavefunction  as a starting point for
understanding the real universe; for a review and references, see
\ref\smolin{L. Smolin, ``Quantum Gravity With A Positive
Cosmological Constant,'' hep-th/0209079.}. Our discussion here
will make it clear how the Kodama state should be interpreted. For
example, in the Fock space that one can build (see \smolin) in
expanding around the Kodama state, gravitons of one helicity will
have positive energy and those of the opposite helicity will have
negative energy.  

\bigskip\noindent{\it Upside Down Wave Function Of The Harmonic Oscillator}

Consider a simple harmonic oscillator with Hamiltonian
$H=(p^2+x^2)/2$. We have set $\hbar=1$ and normalized the
frequency and mass to be 1.  (Accordingly, when we get back to
gauge theory, we will set $g=1$.) The usual ground state wave
function is $\psi= \exp(-x^2/2)$. It is annihilated by the
annihilation operator $a$, and has energy 1/2.  In expanding
around it, one can make a Fock space of states  $(a^*)^n\psi$, of
energy $n+1/2$.

One could also start with the wave function $\psi'=\exp(+x^2/2)$.
For our present purposes, we will not worry about normalization of
the wavefunction (the Chern-Simons wavefunction of Yang-Mills
theory is just as badly behaved as this upside-down Gaussian). We
will just proceed algebraically. One can easily see that $\psi'$
is annihilated by the creation operator $a^*$ and (therefore) is
an eigenfunction of $H=a^*a+1/2=aa^*-1/2$ with energy $-1/2$.
Starting with $\psi'$, one can build a Fock space of states
$a^n\psi'$, with energy   $-n - 1/2$.  The only thing wrong with
this Fock space, apart from the unnormalizability of the
wavefunctions, is that the energies are negative.

Obviously, one cannot define inner products of the states
$a^n\psi'$ by the usual formula $\langle \psi_1|\psi_2\rangle=\int
dx\overline \psi_1\psi_2$, because the integrals will not
converge.  Might there be some other way to define suitable inner
products?  Let us assume there is some inner product relative to
which $x$ and $p$ are hermitian, and hence $a^*$ and $a$ are
adjoints.  We can always normalize the inner product so that
$\langle \psi'|\psi'\rangle=1$.  Then the norm of the first
``excited'' state $a\psi'$ is $\langle
a\psi'|a\psi'\rangle=\langle\psi'|a^*a\psi'\rangle=-\langle
\psi'|\psi'\rangle=-1$.  We used the fact that $a^*a=aa^*-1$ and
that $a^*\psi'=0$.  Continuing in this way, one finds that the
sign of the norm of $a^n\psi'$ is $(-1)^n$.  Similarly, in all of
the other Fock spaces we consider below which contain negative
energy bosonic excitations, half of the states would have negative
norm.

Now suppose one has two harmonic oscillators, with coordinates
$x,y$ and $H=(p_x^2+p_y^2+x^2+y^2)/2$.  Combining the standard
construction for $x$ with the upside-down wave function for $y$,
we take  the wave function $\exp(-(x^2-y^2)/2)$  for the combined
system.  Clearly its energy is 0, as the ground state energies
cancel between $x$ and $y$.  Starting with this state one can make
a Fock space of states, acting with creation operators in $x$ and
annihilation operators in $y$.  The only unusual property is that
the $y$ excitations have negative energy.

One can make a 45 degree rotation of the $x-y$ plane and then this
wavefunction
becomes $\exp(xy)$, an indefinite Gaussian similar to the Chern-Simons wavefunction.

More generally, suppose one has $s$ harmonic oscillators (for any
positive integer $s$) with coordinates $x_i$ and $H=(\sum_i p_i^2
+ (x,Mx))/2$,  where $M$ is any symmetric positive definite matrix
and $(x,Mx)$ is the corresponding quadratic function of $x$. If
$N$ is any matrix such that $N^2=M$, then \eqn\frero{\psi =
\exp(-(x,Nx)/2)} is an eigenfunction of $H$, the ground state
energy being $\Tr N/2$. If $N$ is the (unique) positive square
root of $M$, then one gets the standard ground state.  In this
case, one can proceed to construct the usual Hilbert space of
excitations with positive energy.  In general, for any square root
$N$, one can construct a Fock space, the only oddity being that
some of the modes have negative energy.

\bigskip\noindent{\it Abelian Gauge Theory In Four Dimensions}

Now let us consider the case of $U(1)$ gauge theory in $3+1$ dimensions.
 For
the moment, we work in Coulomb gauge. The role of $x$ is played by
$A_T$, the transverse part of the vector potential $A$ (thus,
$A_T$ is a divergence-free one-form on $\R^3$).  The matrix $M$ is
$*\, d* d$ where $d$ is the exterior derivative and $*$ is the
Hodge star operator. The positive definite square root of $M$ can
be represented by an integral kernel in $\R^3$.  Taking this to
define the wave function, we get the usual ground state for the
free photons.  The ground state energy is positive and divergent
(requiring the standard subtraction) and the excitations have the
standard positive energies.

Instead, $M$ has an obvious local square root, $N=*\,d$  (or
$-*d$). If one uses this, then $N$ is positive for positive
helicity photons and negative for negative helicity photons.  So
the zero-point energy cancels out, analogous to what happens for
the wave function $\exp(xy)$ that was the toy example above.
Moreover, in expanding around this vacuum, one can construct a
Fock space of states; clearly, the positive helicity photons have
positive energy and the negative helicity photons have negative
energy. If we use $-N$ instead of $N$ in constructing the wave
functions, it is positive helicity photons that have negative
energy.

Explcitly, $ (A,NA)={1\over 2}\int d^3x \epsilon^{ijk}A_i
\partial_j A_k$, so the wavefunctions $\exp\left(\pm(A,NA)\right)$
 predicted by this analysis are precisely the Chern-Simons
wavefunctions $\Psi$ and $\tilde\Psi$. (Our analysis really leads
to the Coulomb gauge wavefunctions $\exp(\pm(A_T,NA_T))$, but as
the gauge-invariant generalization of this is merely
$\exp(\pm(A,NA))$, there is no problem in expressing our result in
a gauge-invariant language, and we have done so.)

Since CPT exchanges positive and negative helicities
while commuting with the energy, these results imply that CPT must exchange $\Psi$ with $\tilde\Psi$,
as one can indeed verify directly.  CPT acts by complex conjugation, which leaves both $\Psi$ and $\tilde \Psi$
invariant, combined with a reflection of space, which reverses the sign of the Chern-Simons functional and
so exchanges $\Psi$ and $\tilde\Psi$.

\bigskip\noindent{\it Nonabelian Gauge Theory}

We have now understood the existence of the Chern-Simons wavefunction for abelian gauge theory,
as well as its physical interpretation.  What about the nonabelian case?
The explicit computation that we reviewed at the beginning of this paper showed that the Chern-Simons
wavefunction of Yang-Mills theory
has a simple extension to the nonabelian case.  This computation was so simple
that it is hard to simplify it further,
but I want to explain from a different point of view ``why'' the Chern-Simons
state of nonabelian gauge theory exists.

In general, consider a classical mechanical system with phase
space ${\cal M}$, and with a Lagrangian submanifold ${\cal N}$. At
least formally, one can always associate with ${\cal N}$ a quantum
state $\Psi_{\cal N}$: $\Psi_{\cal N}$ is the state annihilated by
all operators obtained by quantization of functions that vanish on
${\cal N}$.  To see how this works, consider a classical system
with canonical variables $p_i$ and $x^i$, $i=1,\dots,s$. Define a
Lagrangian submanifold ${\cal N}$ by the equations
\eqn\defman{p_i={\partial F\over \partial x^i},} for any function
$F(x^1,\dots,x^s)$.  The corresponding quantum state should be
annihiliated by $p_i-\partial F/\partial x^i$, and, in a
representation in which the $x^i$ act by multiplication and
$p_i=-i\partial/\partial x^i$, it is clearly $\Psi_{\cal
N}=\exp(iF)$.

As this example shows, if ${\cal N}$ is a real Lagrangian
submanifold, then $\Psi_{\cal N}$ is an oscillatory state (and is
normalizable or delta-function normalizable depending on the
global behavior of ${\cal N}$). If one is willing to proceed more
formally, one can replace ${\cal M}$ by its complexification
${\cal M}_{\bf C}$ and let ${\cal N}$ be a complex Lagrangian
submanifold of ${\cal M}_{\bf C}$.  The wavefunction is then a
holomorphic function of the (complexified) coordinates.  The same
formal discussion applies, though the considerations of
normalizability may be quite different.

For example, let us go back to the case of the simple harmonic
oscillator, with phase space variables $x$ and $p$. In the case of
a two-dimensional phase space ${\cal M}$, any codimension-one
submanifold is Lagrangian. So (upon complexification), we can
define a Lagrangian submanifold by $p=ix$, or in other words $p=
dF/dx$ with $F=ix^2/2$. The wavefunction $\Psi_{\cal N}=\exp(iF)$
is then the conventional harmonic oscillator ground state
$\exp(-x^2/2)$.  Alternatively, we could use the Lagrangian
submanifold $p=-ix$, and then we get the highly unnormalizable
wavefunction $\exp(x^2/2)$ that we considered as a step to
explaining the abelian Chern-Simons state. In general, as long as
we work formally and do not worry about normalizability, any
complex Lagrangian submanifold of the complexified phase space
will do.

Now consider in this spirit nonabelian Yang-Mills theory in four
dimensions.  The phase space ${\cal M}$ is the space of classical
solutions of the Yang-Mills equations $D_\mu F^{\mu\nu}=0$ (with
reasonable behavior at spatial infinity), modulo gauge
transformations. In Minkowski space, a non-trivial solution of the
self-dual or anti-self-dual solutions cannot be real.  So we
cannot define a Lagrangian submanifold of ${\cal M}$ by taking
self-dual or anti-self-dual solutions.  Let us, however,
complexify ${\cal M}$.  The complexified space ${\cal M}_C$ is the
space of complex-valued solutions of the Yang-Mills equations (or
if you wish, solutions for a connections that takes values in the
complexification of the Lie algebra).  Since our considerations
are somewhat formal, we do not need to worry about precise
existence theorems for ${\cal M}_{\bf C}$ in what follows.  In
${\cal M}_{\bf C}$, self-dual or anti-self-dual solutions do
exist. To get an anti-self-dual solution, we simply work in the
gauge $A_0=0$ and solve the evolution equation $\partial
A_i/\partial x^0=-(i/2)\epsilon^{ijk}F_{jk}$.  So a solution
exists for arbitrary initial values of $A_i$ at time zero.

The space ${\cal N}$ of complex-valued anti-self-dual solutions is
in fact a Lagrangian submanifold of ${\cal M}_{\bf C}$.  To prove
this, the main point is to show that the symplectic structure
$\omega$ of ${\cal M}_{\bf C}$ vanishes when restricted to ${\cal
N}$.  For this, we use the covariant approach to the canonical
formalism, as described for example in \ref\crnk{C. Crnkovic and
E. Witten, ``Covariant Approach To Canonical Formalism In
Geometrical Theories,'' in S. W. Hawking and W. Israel, eds., {\it
Three Hundred Years Of Gravitation} (Cambridge University
Press).}.  In the canonical formalism, we let $\delta A$ denote a
variation in a classical solution $A$; we treat it as an
anticommuting variable, representing a one-form on the space of
solutions. We then define the symplectic current $J_\mu=\Tr \delta
A^\nu (D_\mu\delta A_\nu-D_\nu\delta A_\mu)$. It is easily shown
to be conserved, and its integral over an arbitrary initial value
hypersurface gives the symplectic two-form $\omega$.  For example,
if we pick the initial value surface to be at $x^0=0$ and work in
the gauge $\delta A_0=0$, we get a formula for $\omega$:
\eqn\omegaq{\omega=\int d^3x\,\,\Tr \delta
A_i{\partial\over\partial x^0}\delta A_i.} Now restricting to
${\cal N}$ means taking $\delta A_i$ to obey
\eqn\nomegaq{{\partial \delta A_i\over \partial
x^0}=-i\epsilon^{ijk}D_j\delta A_k,} which is the linearization of
the anti-self-dual equations.  When we do this, we get
$\omega=-i\int d^3x \epsilon^{ijk}\delta A_i D_j \delta A_k$, and
this vanishes using integration by parts and Fermi statistics for
$\delta A$.\foot{To complete the proof that ${\cal N}$ is
Lagrangian, one needs to show that it is a maximal subspace on
which $\omega$ vanishes. One simply uses the same formulas to show
that if $\delta A=\delta_1A+\delta_2A$, with $\delta_1A$ obeying
the linearization of the self-dual Yang-Mills equations, then
vanishing of $\omega(\delta A)$ for any $\delta_1A$ implies that
$\delta_2A$ obeys the same equation.  So $\omega$ would not vanish
on any enlargement of ${\cal N}$.}

So a quantum state associated with the symplectic manifold ${\cal
N}$ should exist; it should be annihilated by $F^+$, the self-dual
part of $F$.  This quantum state is simply the Chern-Simons
wavefunction $\Psi$.  Indeed, we already showed in \urno\ that
$\Psi$ is annihilated by $F^+$ at time zero.  It follows from this
that $\Psi$ is Poincar\'e invariant (that is, invariant under the
connected part of the Poincar\'e group, though not, as we saw
earlier, under CPT!) and hence is annihilated by $F^+$ at all
times.  To prove Poincar\'e invariance of $\Psi$, note that the
stress tensor $T_{\mu\nu}=\Tr\left(
F_{\mu\alpha}F_\nu{}^\alpha-{1\over 4}
g_{\mu\nu}F_{\alpha\beta}F^{\alpha\beta}\right)$ of Yang-Mills
theory transforms with spin $(1,1)$ under the Lorentz group, while
$F^-$ and $F^+$ transform as $(1,0)$ and $(0,1)$, respectively. So
$T\sim F^+F^-$, and hence any state annihilated by $F^+$ at time
zero is also annihilated at time zero by all components  of
$T_{\mu\nu}$. Hence (as the Poincar\'e generators are certain
integrals of components of $T_{\mu\nu}$ at time zero), such a
state is automatically Poincar\'e invariant.  Poincar\'e
invariance implies that the state is annihilated by $F^+$ at all
times, given that this is the case at time zero.

The covariance is illustrated by the dispersion relation that we
found in the abelian theory (or equivalently in the weak coupling
limit of a nonabelian theory).  The dispersion relation can be
written  $E=\epsilon |\vec p|$ where $E$ is the energy, $\vec p$
the three-momentum, and $\epsilon=\pm 1$ is the sign of the
helicity; this relation is covariant, though exotic.  This form of
the dispersion relation will be preserved when higher order
corrections are considered (to the extent that they make sense
given the unnormalizable ground state and negative energy
excitations), because it is protected by Poincar\'e symmetry.

\bigskip\noindent{\it Supersymmetric Extension}

Now we will, finally, consider the supersymmetric extension of the Chern-Simons state.
Yang-Mills theory can be supersymmetrized, with ${\cal N}=1$ supersymmetry, by simply adding
a Weyl fermion field $\lambda $ that has positive chirality and transforms in the adjoint representation
of the gauge group.  It thus transforms with spin $(0,1/2)$ under Lorentz transformations.
The adjoint field $\bar\lambda$ transforms with spin $(1/2,0)$.
The $\lambda$-dependent part of the Lagrangian is simply the minimally coupled Dirac action
$\int d^4x\, \bar\lambda i\Gamma\cdot D \lambda$.

Classically, there is  a $U(1)$ charge (called in this context an
$R$-symmetry) under which $\lambda$ has charge 1 and $\bar
\lambda$ has charge $-1$.  When the quantum theory is quantized in
the usual way, there is an anomaly in the $R$-symmetry. From a
Hamiltonian point of view, as explained for example in \jackiw,
the anomaly means that homotopically nontrivial gauge
transformations do not commute with the $R$-symmetry. In the
present context, we can ignore this issue because we are anyway
not dividing by large gauge transformations (as discussed in
connection with \urno, we cannot divide by them as the
Chern-Simons state is not invariant under them).

In a fixed gauge field background,
the  fermion state of maximum $R$-charge is the state $\chi$ that is annihilated by all components
of $\lambda$, of either positive or negative frequency; it is not annihilated, therefore, by any
components of $\bar\lambda$.  Of course, there is also a conjugate state $\tilde \chi$ of minimum
$R$-charge, annihilated by all components of $\bar\lambda$.  The states $\chi$ and $\tilde \chi$
are automatically eigenstates of the Hamiltonian, since they are the unique states of their $R$-charge.

Of course, Dirac taught that the proper quantization of this
theory is to fill the Dirac sea and find a state whose excitations
all have positive energy.  This Dirac state can be obtained from
$\chi$ by filling the negative energy states created by half the
modes of $\bar\lambda$, or from $\tilde\chi$ by filling the
negative energy states created by half the modes of $\lambda$. For
our purposes here, however, instead of studying the standard
quantization with the Dirac state, we want to consider the naive
quantization using $\chi$ or $\tilde \chi$.

In expanding around $\chi$, all excitations are created by
components of $\bar\lambda$.  The positive helicity excitations
have positive energy, and the negative helicity excitations have
negative energy.  This is so for a simple and essentially familiar
reason that one can readily understand by recalling the single
particle massless Dirac equation obeyed by the two-component
spinor $\bar\lambda$. This equation reads
\eqn\toggo{i{\partial\over\partial x^0}\bar\lambda =i\vec\sigma
\cdot \vec \nabla \bar\lambda,} or in momentum space
$E=\vec\sigma\cdot \vec p$, where $\vec\sigma$ are $2\times 2$
Pauli matrices, $E$ is the energy, and $\vec p$ is the spatial
momentum. Since $\vec\sigma\cdot\vec p=\epsilon |\vec p|$, where
$\epsilon$ is the sign of the helicity, we get the same dispersion
relation  $E=\epsilon |\vec p|$ that we found in studying the
Chern-Simons state.  (In expanding around the Dirac state, the
negative energy modes created by $\bar\lambda$ are replaced by
modes of positive energy but still negative helicity created by
$\lambda$.  If a negative energy particle has momentum $p$,
angular momentum $J$, and helicity $\epsilon$, then a positive
energy hole representing absence of this particle has momentum
$-p$, angular momentum $-J$, and helicity $\epsilon$.)

This suggests that the naive fermion vacuum $\chi$ is related by
supersymmetry to the Chern-Simons state $\Psi $ for gauge bosons.
This can be seen directly. As the Chern-Simons state is
annihilated by the $(0,1)$ part of $F$, its supersymmetric
extension should be annihilated by the $(0,1/2)$ field $\lambda$,
which is related to $F^+$ by supersymmetry.  Thus, the
supersymmetric extension of the Chern-Simons state should be
annihilated by $\lambda$ as well as $F^+$.   The fermionic part of
such a state is simply $\chi$. So the supersymmetric extension of
the Chern-Simons state is essentially $\Psi\otimes \chi$ (or
$\tilde\Psi\otimes \tilde\chi$ for the state with helicities
reversed).\foot{There is actually a fiber bundle structure here,
rather than a simple tensor product, as the definition of $\chi$
depends on the connection $A$. For our present purposes we ignore
this.} Like the conventional supersymmetric vacuum, the state
$\Psi\otimes \chi$ has zero energy because of supersymmetry. For
the conventional vacuum, the vanishing of the zero-point
contribution to the energy is obtained by a cancellation between
bosons and fermions, while for the state $\Psi\otimes\chi$ the
vanishing is ensured by a cancellation between states (either
bosons or fermions) of positive helicity and states of negative
helicity.

\bigskip
This work was supported in part by NSF Grant PHY-0070928.  While
writing this note, I became aware that some of the results were
also obtained long ago by R. Jackiw. \listrefs
\end